\documentclass[conference,10pt]{IEEEtran}
\usepackage{epsfig,rotating,setspace,latexsym,amsmath,epsf,amssymb,amsfonts,bm,theorem,subfigure,epstopdf}
\usepackage{cite,authblk}
\usepackage{bbm}
\usepackage[ruled,vlined]{algorithm2e}
\usepackage{color}

\SetArgSty{textnormal}

\IEEEoverridecommandlockouts
\allowdisplaybreaks

\newcommand{\rev}[1]{{\color{black}#1}} 

\begin{document}
\bstctlcite{IEEEexample:BSTcontrol} 

\title{Adversarial Attacks against Deep Learning Based Power Control in Wireless Communications}
	
\author[1]{Brian Kim}
\author[2]{Yi Shi}
\author[2]{Yalin E. Sagduyu}
\author[2]{Tugba Erpek}
\author[1]{Sennur Ulukus}

\affil[1]{\normalsize Department of Electrical and Computer Engineering, University of Maryland, College Park, MD 20742, USA}
\affil[2]{\normalsize Intelligent Automation, Inc., Rockville, MD 20855, USA  \thanks{This effort is supported by the U.S. Army Research Office under contract W911NF-20-C-0055. The content of the information does not necessarily reflect the position or the policy of the U.S. Government, and no official endorsement should be inferred.}}
	
\maketitle
\begin{abstract}
We consider adversarial machine learning based attacks on power allocation where the base station (BS) allocates its transmit power to multiple orthogonal subcarriers by using a deep neural network (DNN) to serve multiple user equipments (UEs). The DNN that corresponds to a regression model is trained with channel gains as the input and returns transmit powers as the output. While the BS allocates the transmit powers to the UEs to maximize rates for all UEs, there is an adversary that aims to minimize these rates. The adversary may be an external transmitter that aims to manipulate the inputs to the DNN by interfering with the pilot signals that are transmitted to measure the channel gain. Alternatively, the adversary may be a rogue UE that transmits fabricated channel estimates to the BS. In both cases, the adversary carefully crafts adversarial perturbations to manipulate the inputs to the DNN of the BS subject to an upper bound on the strengths of these perturbations. We consider the attacks targeted on a single UE or all UEs. We compare these attacks with a benchmark, where the adversary scales down the input to the DNN. We show that the adversarial attacks are much more effective than the benchmark attack in terms of reducing the rate of communications. We also show that adversarial attacks are robust to the uncertainty at the adversary including the erroneous knowledge of channel gains and the potential errors in exercising the attacks exactly as specified.
\end{abstract}

\section{Introduction}\label{sec:Introduction}
The algorithmic and computational advances in deep learning (DL) have supported deep neural networks (DNNs) in solving complex problems for various applications such as computer vision \cite{vision1} and speech recognition \cite{speech1}, by effectively learning from large and rich data representations. By capturing the intrinsic characteristics of the spectrum data, DL has been also effectively applied for various wireless communication tasks, such as waveform design, signal classification, spectrum sensing, and interference management \cite{erpek1}.

One particular application of DL in the wireless domain is the transmit power allocation that finds important applications in wireless communications. As the underlying optimization is a complex problem that cannot be readily solved by analytical methods, a data-driven DL approach is largely needed. In this context, the base station (BS) can determine the transmit power allocated to each user equipment (UE) by using a DNN that takes the channel information as the input.  The power allocation with a DNN that takes the position information of UEs as the input has been shown in \cite{Debbah} to achieve near-optimal performance while reducing the complexity which makes it possible to perform power allocation in real time. Furthermore, DL-based power allocation has been also studied for distributed antenna systems \cite{distripower} and cell-free massive multiple-input multiple-output (MIMO) systems \cite{cellfreepower}.

However, the DNNs are highly vulnerable to carefully generated adversarial perturbations that may cause incorrect output or misclassification, as first demonstrated in computer vision applications \cite{Szegedy1}. Moreover, since the wireless medium is shared and open to adversaries such as jammers, the adversarial attack poses a practical threat to the DNNs used in wireless communications. Therefore, adversarial machine learning has recently gained attention in the wireless security domain \cite{Sagduyu2020, adesina2020adversarial}. The attacks built upon adversarial machine learning include exploratory (inference) attacks \cite{Shi2018AdDL4CogRaSec, Terpek}, adversarial (evasion) attacks \cite{Larsson2, Kokalj2, Kokalj3, Flowers1, Bair1, Lin, Kim1, Kim2, Gunduz2, Gunduz1, Kim5G, KimMultiple, KimICC, IoT,Larsson3, sagduyubc}, poisoning (causative) attacks \cite{YiMilcom2018,Sagduyu1, Luo2019, ZluoPartialAttack, Zluo2021}, membership inference attacks \cite{MIA, MIA2}, Trojan attacks \cite{Davaslioglu1}, and spoofing attacks \cite{Shi2019generative, ShiGANSpoofing}. These attacks are stealthier than conventional jamming schemes \cite{Sagduyu2008, Sagduyuuncertainty}. Most of the applications of adversarial attacks to the wireless communications have focused on wireless signal classifiers such as modulation classifiers \cite{Larsson2, Kokalj2, Kokalj3, Flowers1, Bair1, Lin, Kim1, Kim2, Gunduz2, Gunduz1, Kim5G, KimMultiple, KimICC} and spectrum sensing classifiers \cite{IoT,YiMilcom2018,Sagduyu1, sagduyubc}. Adversarial perturbations have been also extended to other communication problems such as autoencoder-based end-to-end communications \cite{Larsson1} and beam prediction \cite{KimSSP}.

In this paper, we use a regression-based DNN at the BS to allocate the power to orthogonal subcarriers and serve multiple users. 
Adversarial attacks on the MIMO power control have been considered in \cite{Larsson3} with the goal of preventing the underlying DNN (that is trained to maximize the product of signal-to-noise-ratios (SNRs) by taking the UE positions as the input) from finding a feasible solution. In this paper, we formulate the power allocation under the attack to rely on a robust and practical DNN solution that always finds a feasible solution for any set of channel estimates given as the input. To launch an attack on this DNN, we consider an adversary that manipulates the input (channel gains) to the DNN in test time to minimize the minimum rate among all UEs. The adversary can be modeled in two ways: (i) the adversary is an external transmitter that aims to manipulate the inputs to the DNN over the air by interfering with the pilot signals that are transmitted to estimate the channel gain, or (ii) the adversary is a rogue UE that transmits fabricated channel estimates back to the BS.

We design an adversarial attack to change the DNN's input to manipulate the minimum rate over all UEs subject to the condition that the perturbations to the inputs of the DNN are bounded. In particular, the adversary generates the adversarial attacks to manipulate the minimum rate by crafting the perturbations to the DNN input based on the gradient of the minimum rate. For that purpose, we consider two approaches to compute the gradient of the rate with respect to the inputs to the DNN when crafting the perturbation: (i) the adversary obtains the DNN power allocation outputs from its surrogate model, computes the minimum rate based on these outputs using analytical means, and then computes the gradient of the rate with respect to the changes to the DNN input, and (ii) the adversary aims to attack the DNN by calculating the gradient of the DNN's loss function using the fast gradient method (FGM), where we define the DNN's loss function as the error with respect to the minimum rate.

We consider the attacks targeted on a single UE or all UEs, i.e., the adversary aims to manipulate the channel gain estimates of a single UE or all UEs, respectively. We compare these attacks with a benchmark attack, namely a scaling down attack, where the adversary scales down the input to the DNN. Our results show that the adversarial attacks are much more effective than the benchmark attack in terms of reducing the rate of communications even if a small perturbation is used. We also show that the adversarial attacks can be effectively launched even under two types of uncertainty at the adversary, (i) the knowledge of channel gains at the adversary is erroneous, and (ii) the adversary cannot generate the exact planned perturbation in the channel gain estimate of the UE. Overall, these results show that the adversarial attacks pose a serious threat to power allocation solutions that rely on deep learning.


The rest of the paper is organized as follows. Section~\ref{sec:victim} provides the system model. Section~\ref{sec:attack} describes the adversarial attacks considered in this paper. Section~\ref{sec:simulation} presents the attack results. Section~\ref{sec:Conclusion} concludes the paper.

\section{Victim Model: DL for Power Allocation} \label{sec:victim}
We consider the power allocation problem for downlink communications from the BS using $N$ different orthogonal subcarriers to communicate with $K$ UEs, where the downlink signal transmitted by the $i$th subcarrier of the BS to the $j$th UE is $x_{ij}$ and the corresponding power is $|x_{ij}|^2 = p_{ij}$. 
The channel between the $i$th subcarrier of the BS to the $j$th UE is $h_{ij}$ and the corresponding channel gain is $g_{ij} = |h_{ij}|^2$. 
\begin{figure}[t]
	\centerline{\includegraphics[width=0.8\linewidth]{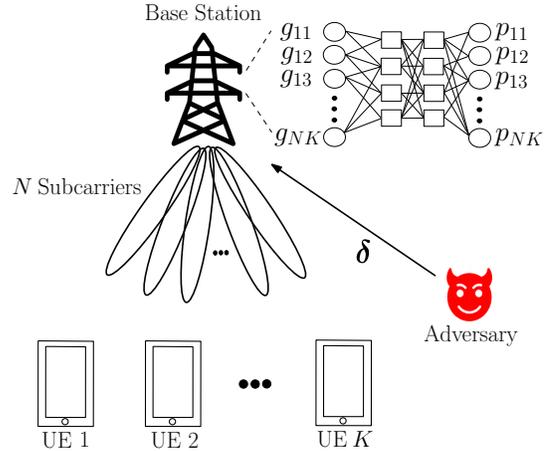}}
	\caption{System model.}
	\label{fig:sys}
\end{figure}
For channel estimation, the BS transmits pilot signals from each of its subcarriers one by one, the UE estimates the channel gains, and reports them back to the BS. Based on these channel estimates, the BS allocates power to its subcarriers to serve each of the UEs. In particular, power $p_{ij}$ for UE $j$’s data at subcarrier $i$ is an optimization variable to be determined by the BS, where $\sum_{i}\sum_{j}p_{ij}\le p$. 
Then the received signal at the $j$th UE \rev{for subcarrier $i$} is 
\begin{equation}\label{eq:received signal}
    s_{j,i} = h_{ij}x_{ij}+  \sum_{k\ne j} h_{ij}x_{ik} + n_i,
\end{equation}
where $n_i$ is the noise with power $\sigma_i^2$. The rate of UE $j$ is given as in \cite{rate_expression} by
\begin{equation}\label{eq:rate}
    r_{j}(\boldsymbol{p}) = \rev{\sum_{i=1}^{N}\log_{2}\left( 1+\frac{g_{ij}p_{ij}}{\sigma_i^2+\sum_{k\ne j}g_{ij}p_{ik}}\right),}
\end{equation}
where $\boldsymbol{p} = [p_{11}, p_{12}, \cdots, p_{NK}]$.

The achievable rates for the UEs are considered by the BS to allocate the transmit power where the objective of the BS can be maximizing the minimum rate of all UEs, namely maximizing $r_{\min}$, where $r_{\min} \leq r_j(\boldsymbol{p})$ for all $j$, by allocating the transmit power to subcarriers for UEs. Therefore, we have the following optimization problem:
\begin{align} \label{eq:power allocation}
	\max_{\boldsymbol{p}}& \quad r_{\min}\nonumber\\
	\mbox{s.t.} & \quad r_{\min}\le r_{j}(\boldsymbol{p}), \quad 1 \le j \le K\nonumber \\
	&\quad \sum_{i=1}^{N} \sum_{j=1}^{K} p_{ij} \le p \nonumber\\
	&\quad p_{ij} \ge 0, \quad 1 \le i \le N, 1 \le j \le K.
\end{align}

This is a nonlinear optimization problem. Although some methods such as interior point and trust region can be applied to solve such nonlinear optimization problems, the complexity could be high for online power allocation under dynamic channel gains. Thus, the BS can build a DL algorithm, namely train a DNN, to solve (\ref{eq:power allocation}). The input for this multi-output regression problem is the set of channel gains $g_{ij}$ (note that $\sigma_i^2$ and $p$ are constants) and the output is the power allocation $p_{ij}$, where the training data (input and output) samples are obtained by solving (\ref{eq:power allocation}) offline. Specifically, we define the dataset,  $\{\boldsymbol{x}(n),\boldsymbol{y}(n)\}^{N_{t}}_{n=1}$, where the input $\boldsymbol{x}(n)$ is the channel gain, the output $\boldsymbol{y}(n)$ is the power allocation, and $N_t$ is the size of the training dataset. We denote the regression-based DNN at the BS as $f(\boldsymbol{x}(n);\boldsymbol{\theta})$, where $\boldsymbol{\theta}$ is the set of DNN parameters, the loss function of the DNN at the BS is $L_g(\boldsymbol{\theta},\boldsymbol{x}(n),\boldsymbol{y}(n))$, and the predicted power allocation $f(\boldsymbol{x}(n),\boldsymbol{\theta})$ is $\hat{\boldsymbol{p}}$.

While the BS determines the power allocation variable $\hat{\boldsymbol{p}}$ from the estimated channel gains using the pre-trained regression-based DNN, there exists an adversary that aims to decrease the minimum rate among UEs by manipulating the channel information at the BS so that the BS makes wrong decisions. While doing so, the adversary may attack one UE (and change its channel gains) or attack all UEs. We assume that there is a budget on these changes, which is measured as the percentage of the total original channel gains. For example, suppose the budget is $1$\% (or, $0.01$). Then, the total change can be no more than $0.01 \sum_i g_{ij}$, if the adversary attacks UE $j$, or $0.01 \sum_i \sum_j g_{ij}$, if the adversary attacks all UEs. The adversary can train its own DNN and use this DNN as a surrogate model for the DNN of the BS. We introduce two approaches to generate an adversarial attack at the adversary, based on this surrogate model. 

\section{Adversarial Attacks on Power Control}\label{sec:attack}

\subsection{Simplified Analytical Gradient-Based Attack}\label{sec:gradient attack}

To maximize the impact of changes at the BS, the adversary needs to carefully spend the budget of changes on channel gains. The first approach that we consider for this purpose is based on the analysis of each channel gain's gradient using (\ref{eq:rate}) and the adversary DNN's power outputs. Denote the gradient for the $i$th subcarrier and UE $j$ as $\eta_{ij}$, which can be determined by channel gains and power values as 
\begin{align} \label{eq:gradient}
\eta_{ij} =
\frac{p_{ij} \sigma_i^2}{(\sigma_i^2+g_{ij}\sum_{k\ne j}p_{ik}) (\sigma_i^2+g_{ij}\sum_{k=1}^N p_{ik}) \log_e 2},
\end{align}
where we simplify the problem by ignoring the dependency between $p_{ij}$ and $g_{ij}$. In Section~\ref{sec:dnn attack}, we will use the gradient through the loss function defined for the DNN to consider this dependency.
Since $\eta_{ij} > 0$, the channel gain from the $i$th subcarrier to UE $j$ should be decreased so that the rate can be smaller. 
Moreover, it is more effective if the adversary changes a channel gain with a larger value $\eta_{ij}$. Thus, 
the adversary first selects the channel gain $g_{ij}$ with the largest $\eta_{ij}$ and tries to decrease this channel gain. 
The details to attack a specified UE $j$ are presented in Algorithm~\ref{alg:change}. 


If the adversary can attack all UEs, it first selects the UE $j$ with the largest $\sum_i \eta_{ij}$. If the budget for change permits, the adversary decreases $g_{ij}, i=1,\cdots, N$, to $0$ and then selects the next UE to attack. Otherwise, the adversary applies Algorithm~\ref{alg:change} to attack this UE. 

\begin{algorithm}[t]
	\DontPrintSemicolon
	\SetAlgoLined
	\label{alg:change}
    \textbf{Input}: the target UE $j$, channel gain $g_{ik}$ for each subcarrier $i$ and UE $k$, budget for total change $B_{g}$, a small threshold $\varepsilon$ for minimum channel gain \\
    \textbf{Calculate}: gradient $\eta_{ij}$ for each subcarrier $i$ by (\ref{eq:gradient})\\
	\textbf{Sort}: $\eta_{ij}$ in a list $A$ based on the non-increasing order\\
	\For{$i \in A$}{
	\lIf{$g_{ij} \ge B_g + \varepsilon$}{$\delta_{ij} = - B_g$ \textbf{break}}
	\lElse {$B_g = B_g - g_{ij} + \varepsilon$ and $\delta_{ij} = \varepsilon - g_{ij}$}
	}
	Output: $\delta_{ij}$ for $i=1,\cdots, N$ 
	\caption{Simplified analytical gradient-based attack algorithm for changing channel gains.}
\end{algorithm}


Once the channel gains are changed by the adversary, the BS makes its decision on power allocation based on the incorrect channel gains and determines the transmitted data rate for each UE based on allocated powers and incorrect channel gains. On the other hand, the maximum link rate for each UE is determined by the allocated powers and real channel gains. If the transmitted rate is no more than the maximum link rate, the achieved rate is the transmitted rate, otherwise the achieved rate is zero since the transmitted data cannot be decoded by an UE.

\subsection{DNN Gradient-Based Attack}\label{sec:dnn attack}
The second approach for the adversary to craft the adversarial perturbation $\boldsymbol{\delta}$ solves the following optimization problem:
\begin{align} \label{eq:adversary optimization}
	\min_{\boldsymbol{\delta}} \max_{\boldsymbol{p}} & \quad r_{\min}\nonumber\\	\mbox{s.t.} &\quad r_{\min}\le r_{j}'(\boldsymbol{\delta}), \quad 1 \leq j \leq K\nonumber \\
	&\quad \sum_{i=1}^{N} \sum_{j=1}^{K} |{\delta}_{ij}| \le B_{g} \nonumber\\
	&\quad \sum_{i=1}^{N} \sum_{j=1}^{K} p_{ij} \le p \nonumber\\
	&\quad p_{ij} \ge 0,  \quad 1 \le i \le N, 1 \le j \le K,
\end{align}
where  
\begin{equation}
\rev{r_{j}'(\boldsymbol{\delta}) = \sum_{i=1}^{N}\log_{2}\left( 1+\frac{(g_{ij}+\delta_{ij})p_{ij}}{\sigma_i^2+ ((g_{ij}+\delta_{ij})\sum_{k\ne j}p_{ik}) }\right)}
\end{equation}
%
%
and $B_{g}$ is the budget for total change at the adversary.
However, solving (\ref{eq:adversary optimization}) is hard due to nonlinearity. Thus, we use the FGM \cite{Kurakin1} to linearize the loss function $L_g(\boldsymbol{\theta},\boldsymbol{x},\boldsymbol{y})$ of the adversary's DNN in a neighborhood of $\boldsymbol{x}$ and use this linearized function to generate an adversarial attack. 
Since the goal of the adversary is to minimize $r_{\min}$, the adversary defines a loss function $L_a(\boldsymbol{\theta},\boldsymbol{x},\boldsymbol{y})$ that calculates $r_{\min}$. Note that the loss function $L_g(\boldsymbol{\theta},\boldsymbol{x},\boldsymbol{y})$ is used for training and the loss function $L_a(\boldsymbol{\theta},\boldsymbol{x},\boldsymbol{y})$ is used to create an attack. 
Therefore, the adversary uses 
\begin{equation} \label{delta}
\boldsymbol{\delta} = B_{g} \frac{\nabla_{\boldsymbol{x}}L_a(\boldsymbol{\theta},\boldsymbol{x},\boldsymbol{y})}{(||\nabla_{\boldsymbol{x}}L_a(\boldsymbol{\theta},\boldsymbol{x},\boldsymbol{y})||_{1})}
\end{equation}
to attack all UEs. Then, the BS receives $g_{ij}+\delta_{ij}$ for $i=1,\cdots,N$ and $j=1,\cdots, K$. 
Note that $g_{ij}+\delta_{ij}$ can be negative or greater than $1$ depending on the $\delta_{ij}$. For this case, the negative value is changed to zero and the value greater than $1$ is changed to $1$, since a channel gain is always in $[0,1]$. Thus, the budget $B_g$ is not fully used. We can fully utilize the budget $B_g$ in (\ref{delta}) by shifting the perturbation so that $g_{ij}+\delta_{ij}$ stays in $[0,1]$. The perturbation $\boldsymbol{\delta}$ to attack a single UE can be determined similarly by changing $\delta_{ik}=0$ for $k\ne j$. The details are presented in Algorithm \ref{alg:attack 3}. 

\begin{algorithm}[t]
	\DontPrintSemicolon
	\SetAlgoLined
	\label{alg:attack 3}
	\textbf{Input}: channel gain $\boldsymbol{x}$, budget for total change $B_{g}$, and architecture of the DNN\\
	\textbf{Loss function}: use $L_{a}$ to generate perturbation that minimizes the rate among all UEs \\
	\textbf{Calculate}: $\nabla_{\boldsymbol{x}}L_{a}(\boldsymbol{\theta},\boldsymbol{x},\boldsymbol{y})$\\
	\If{ $\min\{\nabla_{\boldsymbol{x}}L_{a}(\boldsymbol{\theta},\boldsymbol{x},\boldsymbol{y})\}<0$}{
		
		$\boldsymbol{\eta} = \nabla_{\boldsymbol{x}}L_{a}(\boldsymbol{\theta},\boldsymbol{x},\boldsymbol{y}) - \min\{\nabla_{\boldsymbol{x}}L_{a}(\boldsymbol{\theta},\boldsymbol{x},\boldsymbol{y})\}\boldsymbol{1}$
	}
	\lElse{$\boldsymbol{\eta} =\nabla_{\boldsymbol{x}}L_{a}(\boldsymbol{\theta},\boldsymbol{x},\boldsymbol{y}) $}
	\textbf{Transmit}: $\boldsymbol{\delta} = B_{g}\frac{\boldsymbol{\eta}}{||\boldsymbol{\eta}||_1}$
	
	\caption{DNN gradient-based attack algorithm for changing channel gains.}
\end{algorithm}

\section{Performance Evaluation}\label{sec:simulation}
We consider a feedforward neural network (FNN) trained as a multi-output regression model for power control. As $N$ is the number of subcarriers and $K$ is the number of UEs, both the input layer and the output layer have size $N \times K$, which corresponds to both the total number of channels (the input of the FNN) and the total number of powers to be allocated (the output of the FNN). To ensure the output power constraint (i.e., the sum of power outputs is less than or equal to $p$), the activation function for the output layer is set as softmax (such that the DNN outputs are summed up to $1$) and the output values are multiplied by $p$ to get power values. To collect training data, we generate $50000$ random instances of channel gains for $N=4$, $N=10$, and $N=20$, where the number of UEs, $K$, is fixed as $3$, and then solve (\ref{eq:power allocation}) by the interior point method in MATLAB to obtain the corresponding power allocation and achieved objective value $r_{\min}$. We train three different DNNs for different number of subcarriers and use half of the generated dataset to train the DNN. The noise power $\sigma_i^2$ is $1/N$ and the total power is $p=10$ during the simulations. The DNN structure for the power allocation is given in Table \ref{table1}.

\begin{table}[t]
	\caption{The DNN architecture for the power allocation.}
	\begin{tabular}{l|l|l}
		Layers&Number of neurons  & Activation function  \\ \hline
		Input & $N \times K$ & - \\ \hline
		Dense 1& 1024 & ReLu \\ \hline
		Dense 2& 1024 & ReLu \\ \hline
		Dense 3& 1024 & ReLu \\ \hline
		Dense 3& 512 & ReLu \\ \hline
		Output&  $N \times K$& Softmax
	\end{tabular}\label{table1}
\end{table}

We use different loss functions for the DNN at the BS. The mean absolute error (MAE) loss function aims to minimize the average absolute error between powers $p_{ij}$ of the training data and powers $\hat p_{ij}$ obtained by the DNN, i.e.,
\begin{equation}
l_{\text{MAE}} = \frac{1}{NK} \sum_i \sum_j |p_{ij} - \hat p_{ij}|.
\end{equation}
 The mean absolute percentage error (MAPE) loss function aims to minimize 
\begin{equation}
l_{\text{MAPE}} = \frac{1}{NK} \sum_i \sum_j \frac{|(p_{ij}+c) - (\hat p_{ij}+c)|}{p_{ij}+c},
\end{equation} where a constant $c=10$ is added to all powers to avoid the divided-by-zero issue. The mean squared logarithmic error (MSLE) loss function aims to minimize 
\begin{equation}
l_{\text{MSLE}} = \frac{1}{NK} \sum_i \sum_j (\log(p_{ij}+1) - \log(\hat p_{ij}+1))^2.
\end{equation}

\begin{figure}[t]
	\centerline{\includegraphics[width=0.95\linewidth]{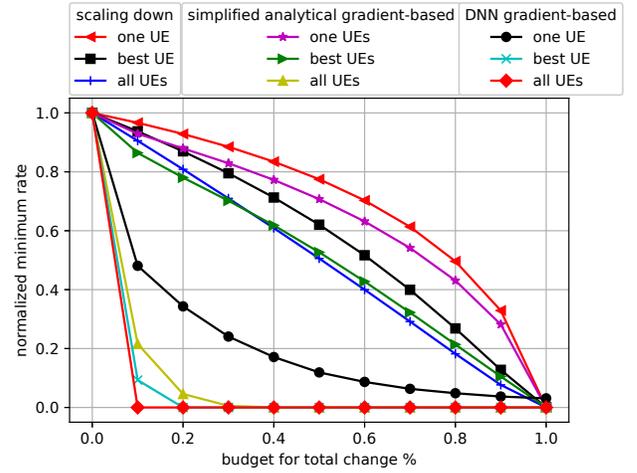}}
	\caption{The normalized minimum rate when $N = 4$ and $K = 3$.}
	\label{fig:4ant}
\end{figure}

We apply the ADAM optimizer and find that DL can always achieve small loss values for all loss functions. However, if we calculate the ratio between the achieved minimum rate using the DNN's output with the minimum rate achieved by training data, these loss functions translate to different performance. For $N = 4$ and $K = 3$, the average ratio between these two rates is $86.37$\% for MSLE, $85.86$\% for MAE, and $84.11$\% for MAPE. Since the aim of the BS is to maximize the minimum rate among all UEs, we define our custom loss function that aims to minimize

\begin{equation}
l_{\text{custom}} =  ( \min_j r_{j}(\boldsymbol{p})-\min_j r_{j}(\hat{\boldsymbol{p}}))^2,
\end{equation}
which achieves $94.45$\% as the ratio between the achieved minimum rate using the DNN's output and the minimum rate achieved by training data when $N = 4$ and $K = 3$. Thus, we adopt the custom loss function as our loss function during the simulations. Note that this is also the loss function that is used to create the adversarial perturbation at the adversary.  Throughout the simulations, we use this normalized rate ratio.
 
In Fig. \ref{fig:4ant}, we compare the two gradient-based attacks when $N = 4$ and $K = 3$. The scaling down attack is also compared as a benchmark attack that enforces the input at the DNN to scale down by $1-\rho$. 
For the one UE case, the adversary always attacks UE $1$. We also consider a hypothetical scheme that the adversary can always find the best UE to attack. The scaling down attack has poor performance compared to the other attacks. 
Attacking all three UEs outperforms the cases where the adversary attacks one fixed UE or the best UE. The DNN gradient-based attack on all UEs outperforms other attack schemes, while the simplified analytical gradient-based attack on all UEs has comparable attack performance. 

\begin{figure}[t]
	\centerline{\includegraphics[width=0.95\linewidth]{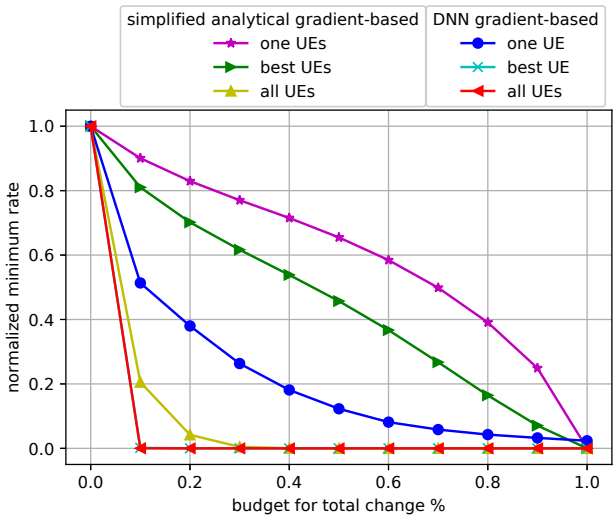}}
	\caption{The normalized minimum rate when $N = 10$ and $K = 3$.}
	\label{fig:10ant}
\end{figure}

\begin{figure}[t]
	\centerline{\includegraphics[width=1\linewidth]{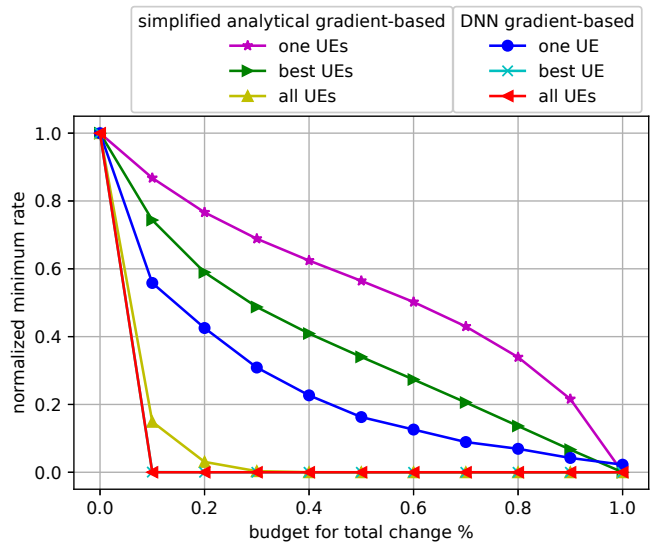}}
	\caption{The normalized minimum rate when $N = 20$ and $K = 3$.}
	\label{fig:20ant}
\end{figure}

Next, we compare different attacks in Fig. \ref{fig:10ant} when $N = 10$ and $K = 3$. 
Without any attack, the regression-based DNN at the BS reaches $90.90$\% ratio between the achieved minimum rate using the DNN's output and the minimum rate achieved by training data. Attacking all UEs simultaneously has more effect on the minimum rate among UEs compared to attacking only one UE for all attack schemes.  
Moreover, the DNN gradient-based attack outperforms the analytical gradient-based attack when all UEs are under attack. It is also observed that when the DNN gradient-based attack is used for attacking the best UE has the same effect as attacking all UEs simultaneously. Similar results are also obtained in Fig. \ref{fig:20ant} when $N = 20$ and $K = 3$. 
Without any attack, the DNN at the BS reaches $83.73$\% ratio between the achieved minimum rate using the DNN's output and the minimum rate achieved by training data.

%

We also consider the impact of uncertainty on adversarial attacks. The first uncertainty is that the adversary may not have accurate knowledge on channel gains, i.e., if the real gain is $g$, the adversary may have an estimate within $[(1-e) g, (1+e) g]$, where $e$ is the error ratio. The second uncertainty is that the adversary may not be able to launch the attack exactly as specified, i.e., if the adversary aims to change a channel gain by amount $c$, the actual change may take a value in $[(1-e) c, (1+e) c]$. This uncertainty may be due to the erroneous knowledge of the adversary about the channel gain from itself to the UE such that it exercises the wrong perturbation power. For the analytical gradient-based attack, Table~\ref{table:uncertainty} shows the normalized minimum rate under uncertainty when the budget to change UE $1$'s channel gains is up to $10$\% when 4 subcarriers are used. Note that when there is no error, the attack leads to $87.77$\% of the rate ratio. Results in Table~\ref{table:uncertainty} show that the errors on channel gains and channel changes will reduce the effect of attack (the rate under attack is higher when there are such errors). However, such effect is small, i.e., the attack is robust to the two types of uncertainty at the adversary.

\begin{table}[t]
	\caption{The impact of uncertainty on attack performance.}
	\centering
	{\small
		\begin{tabular}{c|c|c|c|c}
			Error ratio  & $5$\% & $10$\% & $15$\% & $20$\% \\ \hline \hline
			Error on  & & &  &  \\
			channel gain& $87.77$\% & $87.79$\% & $87.81$\% & $87.84$\% \\ \hline
			Error on  & & &  &  \\
			channel change& $88.02$\% & $88.25$\% & $88.47$\% & $88.67$\%
		\end{tabular}
	}
	\label{table:uncertainty}
\end{table}

\section{Conclusion} \label{sec:Conclusion}
We considered the power control problem at the BS that uses a DNN to maximize the minimum rate among all UEs while the adversary launches adversarial attacks to minimize the minimum rate among all UEs. We considered various methods such as analytical and DNN gradient-based attacks to craft adversarial perturbations on channel estimates targeting one or multiple UEs subject to the budget on adversarial perturbations. Our results showed that the adversarial attacks on power control can significantly reduce the minimum rate among all UEs by slightly manipulating the channel estimate inputs to the DNN of the BS. These attacks remain effective when we vary the number of subcarriers at the BS and when the adversary is subject to errors regarding channel gains and channel changes.

\bibliographystyle{IEEEtran}
\bibliography{lib}

\begin{thebibliography}{10}
\providecommand{\url}[1]{#1}
\csname url@samestyle\endcsname
\providecommand{\newblock}{\relax}
\providecommand{\bibinfo}[2]{#2}
\providecommand{\BIBentrySTDinterwordspacing}{\spaceskip=0pt\relax}
\providecommand{\BIBentryALTinterwordstretchfactor}{4}
\providecommand{\BIBentryALTinterwordspacing}{\spaceskip=\fontdimen2\font plus
\BIBentryALTinterwordstretchfactor\fontdimen3\font minus
  \fontdimen4\font\relax}
\providecommand{\BIBforeignlanguage}[2]{{%
\expandafter\ifx\csname l@#1\endcsname\relax
\typeout{** WARNING: IEEEtran.bst: No hyphenation pattern has been}%
\typeout{** loaded for the language `#1'. Using the pattern for}%
\typeout{** the default language instead.}%
\else
\language=\csname l@#1\endcsname
\fi
#2}}
\providecommand{\BIBdecl}{\relax}
\BIBdecl

\bibitem{vision1}
A.~Krizhevsky, I.~Sutskever, and G.~Hinton, ``Imagenet classification with deep
  convolutional neural networks,'' in \emph{Advances in Neural Information
  Processing Systems (NIPS)}, 2011.

\bibitem{speech1}
G.~Hinton, L.~Deng, D.~Yu, G.~Dahl, A.~Mohamed, A.~Senior, N.~Jaitly,
  V.~Vanhoucke, P.~Nguyen, B.~Kingsburyv, and T.~Sainath, ``Deep neural
  networks for acoustic modeling in speech recognition,'' \emph{IEEE Signal
  Processing Magazine}, vol.~29, pp. 82--97, Nov. 2012.

\bibitem{erpek1}
T.~Erpek, T.~O’Shea, Y.~E. Sagduyu, Y.~Shi, and T.~C. Clancy, ``Deep learning
  for wireless communications,'' in \emph{Development and Analysis of Deep
  Learning Architectures}.\hskip 1em plus 0.5em minus 0.4em\relax Springer,
  Cham, 2020, pp. 223--266.

\bibitem{Debbah}
L.~Sanguinetti, A.~Zappone, and M.~Debbah, ``Deep learning power allocation in
  massive {MIMO},'' in \emph{Asilomar Conference on Signals, Systems, and
  Computers}, 2018.

\bibitem{distripower}
G.~Qian, Z.~Li, C.~He, X.~Li, and X.~Ding, ``Power allocation schemes based on
  deep learning for distributed antenna systems,'' \emph{IEEE Access}, 2020.

\bibitem{cellfreepower}
Y.~Zhao, I.~G. Niemegeers, and S.~H.~D. Groot, ``Power allocation in cell-free
  massive {MIMO}: A deep learning method,'' \emph{IEEE Access}, 2020.

\bibitem{Szegedy1}
C.~Szegedy, W.~Zaremba, I.~Sutskever, J.~Bruna, D.~Erhan, I.~Goodfellow, and
  R.~Fergus, ``Intriguing properties of neural networks,'' \emph{arXiv preprint
  arXiv: 1312.6199}, 2013.

\bibitem{Sagduyu2020}
Y.~E. Sagduyu, Y.~Shi, T.~Erpek, W.~Headley, B.~Flowers, G.~Stantchev, and
  Z.~Lu, ``When wireless security meets machine learning: Motivation,
  challenges, and research directions,'' \emph{arXiv preprint
  arXiv:2001.08883}, 2020.

\bibitem{adesina2020adversarial}
D.~Adesina, C.-C. Hsieh, Y.~E. Sagduyu, and L.~Qian, ``Adversarial machine
  learning in wireless communications using {RF} data: A review,'' \emph{arXiv
  preprint arXiv:2012.14392}, 2020.

\bibitem{Shi2018AdDL4CogRaSec}
Y.~{Shi}, Y.~E. {Sagduyu}, T.~{Erpek}, K.~{Davaslioglu}, Z.~{Lu}, and J.~H.
  {Li}, ``Adversarial deep learning for cognitive radio security: Jamming
  attack and defense strategies,'' in \emph{IEEE International Conference on
  Communications (ICC)}, 2018.

\bibitem{Terpek}
T.~Erpek, Y.~E. Sagduyu, and Y.~Shi, ``Deep learning for launching and
  mitigating wireless jamming attacks,'' \emph{IEEE Transactions on Cognitive
  Communications and Networking}, vol.~5, no.~1, pp. 2--14, Mar. 2019.

\bibitem{Larsson2}
M.~Sadeghi and E.~G. Larsson, ``Adversarial attacks on deep-learning based
  radio signal classification,'' \emph{IEEE Communications Letters}, vol.~8,
  no.~1, pp. 213--216, Feb. 2019.

\bibitem{Kokalj2}
S.~Kokalj-Filipovic and R.~Miller, ``Targeted adversarial examples against {RF}
  deep classifiers,'' in \emph{ACM WiSec Workshop on Wireless Security and
  Machine Learning (WiseML)}, 2019.

\bibitem{Kokalj3}
S.~Kokalj-Filipovic, R.~Miller, and G.~M. Vanhoy, ``Adversarial examples in
  {RF} deep learning: Detection and physical robustness,'' in \emph{IEEE Global
  Conference on Signal and Information Processing (GlobalSIP)}, 2019.

\bibitem{Flowers1}
B.~Flowers, R.~M. Buehrer, and W.~C. Headley, ``Evaluating adversarial evasion
  attacks in the context of wireless communications,'' \emph{arXiv preprint
  arXiv:1903.01563}, 2019.

\bibitem{Bair1}
S.~Bair, M.~Delvecchio, B.~Flowers, A.~J. Michaels, and W.~C. Headley, ``On the
  limitations of targeted adversarial evasion attacks against deep learning
  enabled modulation recognition,'' in \emph{ACM WiSec Workshop on Wireless
  Security and Machine Learning (WiseML)}, 2019.

\bibitem{Lin}
Y.~Lin, H.~Zhao, Y.~Tu, S.~Mao, and Z.~Dou, ``Threats of adversarial attacks in
  {DNN}-based modulation recognition,'' in \emph{International Conference on
  Computer Communications (INFOCOM)}, 2020.

\bibitem{Kim1}
B.~Kim, Y.~E. Sagduyu, K.~Davaslioglu, T.~Erpek, and S.~Ulukus, ``Over-the-air
  adversarial attacks on deep learning based modulation classifier over
  wireless channels,'' in \emph{Conference on Information Sciences and Systems
  (CISS)}, 2020.

\bibitem{Kim2}
B.~Kim, Y.~E. Sagduyu, K.~Davaslioglu, T.~Erpek, and S.~Ulukus, ``Channel-aware
  adversarial attacks against deep learning-based wireless signal
  classifiers,'' \emph{arXiv preprint arXiv:2005.05321}, 2020.

\bibitem{Gunduz2}
M.~Z. Hameed, A.~Gyorgy, and D.~Gunduz, ``Communication without interception:
  Defense against modulation detection,'' in \emph{IEEE Global Conference on
  Signal and Information Processing (GlobalSIP)}, 2019.

\bibitem{Gunduz1}
M.~Z. {Hameed}, A.~{György}, and D.~{Gündüz}, ``The best defense is a good
  offense: Adversarial attacks to avoid modulation detection,'' \emph{IEEE
  Transactions on Information Forensics and Security}, vol.~16, pp. 1074--1087,
  Sep. 2021.

\bibitem{Kim5G}
B.~Kim, Y.~E. Sagduyu, K.~Davaslioglu, T.~Erpek, and S.~Ulukus, ``How to make
  {5G} communications ``invisible'' adversarial machine learning for wireless
  privacy,'' in \emph{Asilomar Conference on Signals, Systems, and Computers},
  2020.

\bibitem{KimMultiple}
B.~Kim, Y.~E. Sagduyu, K.~Davaslioglu, T.~Erpek, and S.~Ulukus, ``Adversarial
  attacks with multiple antennas against deep learning-based modulation
  classifiers,'' in \emph{IEEE Global Communications Conference (Globecom)},
  2020.

\bibitem{KimICC}
B.~Kim, Y.~E. Sagduyu, T.~Erpek, K.~Davaslioglu, and S.~Ulukus, ``Channel
  effects on surrogate models of adversarial attacks against wireless signal
  classifiers,'' in \emph{IEEE International Conference on Communications
  (ICC)}, 2020.

\bibitem{IoT}
Y.~E. {Sagduyu}, Y.~{Shi}, and T.~{Erpek}, ``{IoT} network security from the
  perspective of adversarial deep learning,'' in \emph{IEEE International
  Conference on Sensing, Communication, and Networking}, 2019.

\bibitem{Larsson3}
B.~Manoj, M.~Sadeghi, and E.~G. Larsson, ``Adversarial attacks on deep learning
  based power allocation in a massive {mimo} network,'' \emph{arXiv preprint
  arXiv:2101.12090}, 2021.

\bibitem{sagduyubc}
Y.~E. Sagduyu, T.~Erpek, and Y.~Shi, ``Adversarial machine learning for {5G}
  communications security,'' \emph{arXiv preprint arXiv:2101.02656}, 2021.

\bibitem{YiMilcom2018}
Y.~Shi, T.~Erpek, Y.~E. Sagduyu, and J.~Li, ``Spectrum data poisoning with
  adversarial deep learning,'' in \emph{IEEE Military Communications Conference
  (MILCOM)}, 2018.

\bibitem{Sagduyu1}
Y.~E. Sagduyu, T.~Erpek, and Y.~Shi, ``Adversarial deep learning for
  over-the-air spectrum poisoning attacks,'' \emph{IEEE Transactions on Mobile
  Computing}, vol.~20, no.~2, pp. 306--319, Oct. 2019.

\bibitem{Luo2019}
Z.~Luo, S.~Zhao, Z.~Lu, J.~Xu, and Y.~Sagduyu, ``When attackers meet {AI}:
  Learning-empowered attacks in cooperative spectrum sensing,'' \emph{IEEE
  Transactions on Mobile Computing}, 2020.

\bibitem{ZluoPartialAttack}
Z.~Luo, S.~Zhao, Z.~Lu, Y.~E. Sagduyu, and J.~Xu, ``Adversarial machine
  learning based partial-model attack in {IoT},'' in \emph{ACM Workshop on
  Wireless Security and Machine Learning ({WiseML})}, 2020.

\bibitem{Zluo2021}
Z.~Luo, S.~Zhao, R.~Duan, Z.~Lu, Y.~E. Sagduyu, and J.~Xu, ``Low-cost
  influence-limiting defense against adversarial machine learning attacks in
  cooperative spectrum sensing,'' in \emph{ACM Workshop on Wireless Security
  and Machine Learning ({WiseML})}, 2021.

\bibitem{MIA}
Y.~Shi, K.~Davaslioglu, and Y.~E. Sagduyu, ``Over-the-air membership inference
  attacks as privacy threats for deep learning-based wireless signal
  classifiers,'' in \emph{ACM WiSec Workshop on Wireless Security and Machine
  Learning (WiseML)}, 2020.

\bibitem{MIA2}
Y.~Shi and Y.~E. Sagduyu, ``Membership inference attack and defense for
  wireless signal classifiers with deep learning,'' \emph{arXiv preprint
  arXiv:2107.12173}, 2021.

\bibitem{Davaslioglu1}
K.~Davaslioglu and Y.~E. Sagduyu, ``Trojan attacks on wireless signal
  classification with adversarial machine learning,'' in \emph{IEEE DySPAN
  Workshop on Data-Driven Dynamic Spectrum Sharing}, 2019.

\bibitem{Shi2019generative}
Y.~Shi, K.~Davaslioglu, and Y.~E. Sagduyu, ``Generative adversarial network for
  wireless signal spoofing,'' in \emph{ACM Workshop on Wireless Security and
  Machine Learning (WiseML)}, 2019.

\bibitem{ShiGANSpoofing}
Y.~Shi, K.~Davaslioglu, and Y.~E. Sagduyu, ``Generative adversarial network in
  the air: Deep adversarial learning for wireless signal spoofing,'' \emph{IEEE
  Transactions on Cognitive Communications and Networking}, vol.~7, no.~1, pp.
  294--303, Mar. 2021.

\bibitem{Sagduyu2008}
Y.~E. Sagduyu, R.~Berry, and A.~Ephremides, ``Jamming games in wireless
  networks with incomplete information,'' \emph{IEEE Communications Magazine},
  vol.~49, no.~8, pp. 112--118, Aug. 2008.

\bibitem{Sagduyuuncertainty}
Y.~E. Sagduyu, R.~A. Berry, and A.~Ephremides, ``Wireless jamming attacks under
  dynamic traffic uncertainty,'' in \emph{International Symposium on Modeling
  and Optimization in Mobile, Ad Hoc, and Wireless Networks (WiOpt)}, 2010.

\bibitem{Larsson1}
M.~Sadeghi and E.~G. Larsson, ``Physical adversarial attacks against end-to-end
  autoencoder communication systems,'' \emph{IEEE Communications Letters},
  vol.~23, no.~5, pp. 847--850, May 2019.

\bibitem{KimSSP}
B.~Kim, Y.~E. Sagduyu, T.~Erpek, and S.~Ulukus, ``Adversarial attacks on deep
  learning based {mmWave} beam prediction in 5{G} and beyond,'' in \emph{IEEE
  Statistical Signal Processing Workshop}, 2021.

\bibitem{rate_expression}
Q.~Shi, W.~Xu, D.~Li, Y.~Wang, X.~Gu, and W.~Li, ``On the energy efficiency
  optimality of ofdma for siso-ofdm downlink system,'' \emph{IEEE
  Communications Letters}, vol.~17, no.~3, pp. 541--544, 2013.

\bibitem{Kurakin1}
A.~Kurakin, I.~Goodfellow, and S.~Bengio, ``Adversarial examples in the
  physical world,'' in \emph{International Conference on Learning
  Representations (ICLR)}, 2017.

\end{thebibliography}

\end{document}